# Topology-Controlled Phonon Dielectric Response Beyond Density Scaling in Metal-Organic Frameworks

Debayan Mondal[1], Jiahao Ye[1], Lorenzo Donà[2], and Jin-Chong Tan[1,*]

[1]*Multifunctional Materials and Composites (MMC) Laboratory, Department of Engineering Science, University of Oxford, Parks Road, Oxford OX1 3PJ, United Kingdom*

[2]*Department of Chemistry, NIS and INSTM Reference Centre, University of Turin, Torino 10125, Italy*

[*]*Corresponding author: jin-chong.tan@eng.ox.ac.uk*

Effective-medium theory treats material porosity as passive dilution. Using *ab initio* density functional theory and high-resolution synchrotron terahertz (THz) spectroscopy on isochemical zeolitic-imidazolate frameworks, we show that while the electronic permittivity obeys Clausius–Mossotti density scaling, the phonon contribution violates the conventional density scaling rules. Identical Born charges rule out the role of local chemistry. Instead, the framework connectivity localizes the THz response, where the coherency of phonon eigenvectors determines the mode-effective charges. Long-range architecture of framework topology ubiquitous in metal-organic frameworks is therefore a dielectric degree of freedom beyond the density scaling of conventional solids.

## I. INTRODUCTION

In porous solids, the dielectric response is usually attributed to two quantities: the mass density and the polarizability of the solid phase. Effective-medium theories (EMT) such as the Clausius-Mossotti and Maxwell-Garnett [1-5] employ the above quantities as inputs, treating the void as a vacuum, for determining the permittivity of the framework [6,7]. Local chemistry, density, and porosity have become the standard parameters for tuning the dielectric response in porous solids [8-10]. Underlying this approach is an assumption that is rarely stated: that the way these building blocks are connected over a long-range scale, the framework topology at fixed local chemistry, has negligible impact on dielectric response in the terahertz (THz) or far-infrared spectral range, where phonon response is prevalent in metal-organic frameworks (MOFs) [9,11]. This Letter tests this hypothesis. We interrogate whether long-range MOF topology contributes to the dielectric response beyond what density and local polarizability already being predicted *via* EMT.

Answering the foregoing question cogently requires an isochemical pair of MOF structures, by which we mean two frameworks that share the same chemical building blocks, *i.e.* metal centre, ligand, and coordination geometry but differ only in how those units are connected. Polymorph pairs of this kind are commonly found in zeolitic imidazolate frameworks (ZIF) and isoreticular MOF series [11-13], yet they have not been used for a dielectric comparison study in a comprehensive way. Topology-dependent permittivity has been reported for a ZIF polymorph pair at the device level [14], but without separating the electronic and phonon contributions. Most of the comparison in literature changes chemistry and topology together, so the determined dielectric difference mixes several variables, such that the topological part cannot be separated [8,10]. Our test isolates a single quantity, which we term the residual $\Delta\varepsilon$, representing the dielectric difference between two isochemical frameworks that remains once density scaling has been enforced, and that no continuum EMT could reproduce by incorporating void fraction alone [15]. A residual of this kind would establish if framework topology may act as a control variable for tuning MOF-based dielectrics.

The ZIF-71 and ZIF-72 pair of framework structures [Fig. 1(a)] offers exactly the differential topology we need for this study. Both materials are constructed from the same Zn(II)–4,5-dichloroimidazolate (dcIm) building blocks, where the $ZnN_4$ tetrahedral nodes and dcIm linkers self-assemble to yield either the porous RHO topology in ZIF-71, or the dense LCS topology in ZIF-72 [13,16]. We computed the static and frequency-dependent permittivity of both frameworks from *ab initio* density functional theory (DFT) (coupled-perturbed Hartree–Fock / Kohn–Sham (CPHF/KS) [17] for the dielectric function $\varepsilon(\omega)$ $[0 \rightarrow \infty]$ and Born charges; density-functional perturbation theory (DFPT) [18-22] for the phonon-mediated response, and measured it experimentally by synchrotron infrared specular reflectance at the Diamond Light Source across the broadband infrared regime [23,24] of 30–4000 $cm^{-1}$. Theory and experiments converged on the same result. The electronic contribution obeys density scaling to within 5–7%, just as Clausius–Mossotti predicts. But the phonon contribution does not obey this scaling. It leaves an extra residual part that no continuum mixing rule can reproduce, whether the Maxwell–Garnett, Bruggeman [2], or Hashin–Shtrikman [4,5]. This is because the phonon permittivity depends on how coherently the phonon eigenvectors extend across the extended framework structure, rather than on the local density of polarizable units alone. We trace the residual to a connectivity-driven depolarization mechanism, which we call topological field exclusion, and we confirm it in a second, chemically distinct pair of ZIF-8 topologies, comprising the porous sodalite (SOD) and the dense diamond (dia) polymorphs. Taken together, the two pairs span four topologies and two chemistries and identify porous-versus-dense connectivity as a control variable for phonon-mediated dielectric response. The vibration mechanism operates in the THz (far-infrared) regime that is attributed to the phonon-controlled dielectric response, a spectral range where the chemical and optical descriptors used in most prior studies [25,26] have failed to account for. Long-range framework architecture of MOFs, in this isochemical pair, is therefore an additional control variable for the dielectric response, distinct from the density-driven scaling of conventional solids.

## II. RESULTS AND DISCUSSION

ZIF-71 and ZIF-72 share the same Zn(II)–4,5-dichloroimidazolate (dcIm) building blocks as depicted in Fig. 1(a). The Zn–N bond lengths and dcIm–Zn–dcIm bond angles are similar in both framework structures, so the local chemistry is identical by construction. What differs is how those building blocks are connected at longer length scales. ZIF-71 adopts the porous RHO topology ($Pm\bar{3}m$; solvent-accessible volume 23%; theoretical density 1.12 g cm$^{-3}$). ZIF-72 adopts the dense LCS topology ($Ia\bar{3}d$; 0.1%; 1.77 g cm$^{-3}$) [27] [Fig. 1(a)]. The CPHF/KS isotropic polarizability per $Zn(dcIm)_2$ formula unit is 21.19 Å$^3$ in ZIF-71 and 22.63 Å$^3$ in ZIF-72, equivalent to within 6.4%.

Combining CPHF/KS and DFPT, we found a static dielectric constant $\varepsilon_0$ of 1.80 for ZIF-71 [Fig. 1(b)] and 2.45 for ZIF-72 [Fig. 1(c)], hence a 36% enhancement; see Methods in Supplemental Materials (SM §S1.5). Synchrotron infrared specular reflectance measured at the B22 MIRIAM beamline (Diamond Light Source), was converted to the dielectric function $\varepsilon(\omega)$ by employing the Kramers–Kronig transformation [24,28]. The measured values are $\varepsilon_0^{\mathrm{exp}}$ = 1.67 ± 0.06 for ZIF-71 and 2.30 ± 0.07 for ZIF-72, which are in close agreement with DFT. The question is, which part of $\varepsilon_0$ contributes to the difference? We split it into the electronic contribution $\varepsilon_\infty$ ascribed to the clamped-ion response, and the lattice-polar (phonon) contribution $\Delta\varepsilon_{\mathrm{phon}} \equiv \varepsilon_0 - \varepsilon_\infty$, which is the other dielectric component originating from displacing the nuclei about their equilibrium positions under an applied field (summed over the infrared-active THz modes) [29,30]. We restrict the analysis to these two components because they are the only intrinsic, bulk-crystalline contributions to the frequency-dependent dielectric function $\varepsilon(\omega)$: the dipolar (GHz) and space-charge (Hz–kHz) contributions associated with molecular reorientation and mobile or interfacial charges can be neglected in these insulating frameworks [31,32]. These two parts of $\varepsilon_0$ are affected very differently by the change in topology. The electronic part risen by up to 24% (theory 1.57 → 1.95; experiment 1.48 → 1.77) while the lattice-polar/phonon part more than doubled (theory 0.23 → 0.50; experiment 0.19 → 0.53); see SM §S6.1. Therefore, the topological change affects primarily through the phonon component.

**FIG. 1 (on Page 8)**

The Clausius–Mossotti relationship [4,33] provides the quantitative test for our comparative study. For two materials with the same polarizability per repeat unit, the function $\mathbb{C}(\varepsilon) = \frac{\varepsilon - 1}{\varepsilon + 2}$ should scale in proportion to mass density. Theory gives $\mathbb{C}(\varepsilon_\infty)$ = 0.160 for ZIF-71 and 0.241 for ZIF-72, giving a ratio of 0.664. Experiment gives 0.138 and 0.204, hence a ratio of 0.675. The mass-density ratio is 0.633. Both theory and experimental values match density scaling to within 5–7%. However, for the phonon-mediated response, the ratio of $\Delta\varepsilon_{\mathrm{phon}}(\mathrm{ZIF}-71)$ / $\Delta\varepsilon_{\mathrm{phon}}(\mathrm{ZIF}-72)$ is 0.460 from theory and 0.358 by experiment, thus lying 27% and 43% below the density ratio, respectively (SM Table S3). Clausius–Mossotti alone cannot explain the suppression in either theory or experiment. Henceforth this is the residual we evaluated: a dielectric difference $\Delta\varepsilon$ that persisted despite density scaling and demands a microscopic explanation. The same residual appears in the second isochemical pair (SM §S8-S9), namely the ZIF-8(SOD)/ZIF-8(dia) structures, computed at the identical level of theory. We found the electronic ratio $\mathbb{C}(\varepsilon_\infty)$ agrees with density ratio to within 3%, whereas the ratio from phonon contribution $\Delta\varepsilon_{\mathrm{phon}}$(SOD)/$\Delta\varepsilon_{\mathrm{phon}}$(dia) = 0.48, 19% lower. The residual is therefore applicable across chemistry and topology, not just unique to the ZIF-71(RHO)/ZIF-72(LCS) pair.

More sophisticated continuum theories (SM §S6.3) also cannot overcome the discrepancy observed above. We tested the representative two-phase isotropic mixing over its physical range, treating ZIF-72 as the dielectric host and the RHO cavities of ZIF-71 as vacuum inclusions at volume fraction $\varphi = 0.23$. Maxwell–Garnett (MG) [15] with the standard spherical depolarization factor of $L$ = 1/3, gives $\Delta\varepsilon_{\mathrm{phon}}^{\mathrm{MG}} = 0.323$. Bruggeman symmetric [2] gives 0.351, while the lower and upper bounds of Hashin–Shtrikman [5] yielded 0.323–0.356. All the continuum mixing rules overpredict the phonon contribution $\Delta\varepsilon_{\mathrm{phon}}$ by 40–60% in theory (DFT) and exceeding 100% in comparison with experiments, and the discrepancy persisted across the MG range of $0.20 \leq L \leq 0.50$ (SM Table S4). Continuum mixing theory fails here for a basic reason; it treats the electric field inside each pore as if it were smooth and uniform and then uses one average number to describe it. However, the real dielectric response of MOFs comes from individual vibrations, each adding up differently across the framework. Averaging the field into a single number eliminates that detail, so the mixing theories cannot give the correct answer. Likewise, the same discrepancy is evident in the ZIF-8(SOD)/ZIF-8(dia) pair, where every member of the family overshoots the actual phonon-mediated permittivity by 20–50% (SM §S9, Table S6). However, the gap is smaller than in the ZIF-71/ZIF-72 pair, since the SOD cage of ZIF-8 is much smaller than the RHO cage of ZIF-71, suggesting a direct indication that the residual scales with the cavity size of porous MOFs.

**FIG. 2 (on Page 9)**

The suppressed phonon contribution in ZIF-71 can be directly pinpointed. The electrostatic potential $V(\mathbf{r})$ [Fig. 2(a)] partitions sharply in ZIF-71 but it varies continuously across the ZIF-72's unit cell; the planar average $V(z)$ and field gradient d$E$/d$z$ [Fig. 2(b,c)] also fall to near-zero plateaus over the RHO cavity, while ZIF-72 sustains gradients throughout. This real-space partition is a useful visualization of where the framework decouples from the applied field, but this is not the precise mechanism. In dielectric-continuum terms [33,34], induced surface charges at the cavity-framework boundary produce a depolarization field that opposes the applied field inside the cavity. The linkers lining the cavity therefore experience a reduced local field, and their contribution to the dielectric sum is suppressed [29,35]. We refer to this mechanism as the topological field exclusion. This idea is connected to phonon localization proposed earlier by Bell and Dean, and later by Allen and Feldman [36,37], who showed how vibrations become localized in disordered solids. However, this effect has not been explored in the context of a MOF structure.

If the phonon contribution $\Delta\varepsilon_{\mathrm{phon}}$ were determined by atomic Born-effective-charge magnitudes alone [29,38], the pair of frameworks would respond almost identically. We therefore computed the full Born effective charge tensor $\boldsymbol{Z}^{*}$ at every symmetry-inequivalent sites. The medians of $|Z^{*}{}_{\mathrm{max}}|$ are statistically indistinguishable for all four atomic species ($|Z^{*}_{\mathrm{Zn}}| \approx 2.33$ *e*; $|Z^{*}_{\mathrm{Cl}}| \approx 0.73$ *e*; $|Z^{*}_{\mathrm{N}}| \approx 1.56$ *e*; $|Z^{*}_{\mathrm{C}}| \approx 1.04$ *e*; see Fig. 3(a), where *e* is the elementary charge). A naïve atom-wise estimator that sums $\sum\left|Z^{*}{}_{\mu}\right|^{2}/M_{\mu}$ for each atom $\mu$ of mass $M_{\mu}$, therefore predicts $\Delta\varepsilon_{\mathrm{phon}}(\mathrm{ZIF}-71)$ to within 6% of $\Delta\varepsilon_{\mathrm{phon}}(\mathrm{ZIF}-72)$. The ratio is 0.46 in theory and 0.36 in experiment (SM Table S3). That is a factor-of-two suppression that an atom-wise sum cannot attain. The results on the tensor anisotropy are shown in Fig. 3(b). Zn nodes in ZIF-71 exhibit a markedly broader distribution of $|Z^{*}{}_{\mathrm{max}}| - |Z^{*}{}_{\mathrm{min}}|$ than those in ZIF-72, just as the geometric partitioning of Fig. 2 anticipates. Pore-edge Zn nodes and bulk Zn nodes are chemically identical but spatially inequivalent, and the Born-charge tensor captures that local inhomogeneity.

**FIG. 3 (on Page 10)**

The reason $\Delta\varepsilon_{\mathrm{phon}}$ does not reduce to atomic dipole magnitudes is because it is a sum over the mode-effective charges, not over the individual atoms. For each infrared-active phonon *k* with frequency $\omega_{k}$ and mass-weighted eigenvector $e_{k,\mu}$ on atom $\mu$ of mass $M_{\mu}$, the relevant quantity is the mode-effective Born charge vector [20,29,39]:

$$\tilde{Z}_{k,\alpha} = \sum_{\mu,\beta} Z^{*}_{\mu,\alpha\beta} \frac{e_{k,\mu,\beta}}{\sqrt{M_{\mu}}} \tag{1}$$

which participate in the phonon dielectric sum as $\Delta\varepsilon_{k} \propto \frac{|\tilde{Z}_{k}|^{2}}{{\omega_{k}}^{2}}$. The magnitude of $\tilde{Z}_{k}$ does not depend on the atomic $Z^{*}$ magnitudes alone, but on how the eigenvector aligns for the per-atom contributions. A delocalized mode adds $Z^{*}$-weighted contributions across many linkers that vibrate in the same direction, resulting in a large oscillator strength. A localized mode samples one or two linkers or metal nodes and is bound by a single dipole. The spatial extent (delocalization) of each mode *k* is quantified by its participation ratio (*PR*) [40,41],

$$PR_{k} = \frac{\left(\sum_{\mu} A_{\mu,k}\right)^{2}}{\left[N \sum_{\mu} {A_{\mu,k}}^{2}\right]} \tag{2}$$

where $A_{\mu,k} = |e_{\mu,k}|^2$. $PR = 1/N$ when a single atom moves, $PR = 1$ when all $N$ atoms move together, with $N$ the number of atoms in the primitive cell [42,43]. We restrict the $PR$ analysis employing Eq. (2) to the frequency window of 0–10 THz, where most of the dielectric modes are found [26]; intramolecular modes above 10 THz are intrinsically localized and contribute weakly to $\Delta\varepsilon_{\mathrm{phon}}$ through the $1/\omega^2$ weighting and are nearly identical in the two frameworks. Within this window, the distributions differ sharply [Fig. 3(c,d)]. In ZIF-71, 91% of the modes are localized ($PR < 0.5$) and 2.5% are delocalized ($PR \geq 0.7$) [inset: pie chart, Fig. 3(c)]. In ZIF-72 the localized fraction drops to 61.5%, the intermediate band ($0.5 \leq PR < 0.7$) rises from 6.5% to 26.0%, and the delocalized fraction rises from 2.5% to 12.5% [inset, Fig. 3(d)]. The delocalized modes of ZIF-72 lie below 4 THz, where $1/\omega^2$ maximizes their dielectric contribution.

This mode-resolved contrast also appears in the total dielectric function $\varepsilon(\omega)$ in Fig. 4. The absorption spectra Im[$\varepsilon(\omega)$] of Fig. 4(a) differentiates the two frameworks sharply. ZIF-72 shows strong resonances at 130, 230, and 280 cm$^{-1}$, whereas ZIF-71 displays weaker vibrations, most prominently near 170 cm$^{-1}$. Every major IR-active phonon peak position is reproduced consistently across the DFT predictions and experiments for both materials. Fig. 4(b) shows a decomposition of the mode-resolved dielectric response, where $\sum_k \Delta\varepsilon_k = \Delta\varepsilon_{\mathrm{phon}}$, giving a quantitative comparison between the two structures. Remarkably, the collective modes of ZIF-72 identified near 30, 97, 130, 230, and 280 cm$^{-1}$ account for almost the entire phonon contributions; the 30 cm$^{-1}$ mode exhibits its dominant static contribution through the $1/\omega^2$ weighting despite a modest Im[$\varepsilon(\omega)$] amplitude. The dielectric contribution in ZIF-71 is spread across about twenty weaker modes. Summing $\Delta\varepsilon_k$ over all IR-active phonons recovers the total $\Delta\varepsilon_{\mathrm{phon}}$ in both frameworks to within 1%. The eigenvector patterns illustrate the underlying mechanism. At 230 and 280 cm$^{-1}$, the ZIF-72 modes show Zn-N motions coordinated across many framework nodes, with aligned displacement vectors. The ZIF-71 modes at the same frequencies are localized around a single RHO cavity, where the per-atom dipoles partially cancel [Fig. 4(c,d); SM Movies S1–S4]. The suppression of $\Delta\varepsilon_{\mathrm{phon}}$in ZIF-71 therefore arises not from the smaller atomic Born charges (Fig. 3), but from the localization of eigenvectors that prevents coherent addition through Eq. (1).

**FIG. 4 (on Page 11)**

Finally, to test generality beyond ZIF-71/-72, we repeated the full analysis on second isochemical pair comprising ZIF-8 SOD/dia [Zn(2-methylimidazolate)$_2$] structures. The same electronic-versus-phonon dichotomy holds: the electronic channel obeys the Clausius–Mossotti relation, but the phonon component does not, and the residual scales with cavity size. Details and analyses are presented in SM §S8-S9.

## III. CONCLUSIONS

At fixed local chemistry, density-based continuum theory fails to capture the phonon contribution of the dielectric response of a porous framework material. In the isochemical pair of ZIF-71/ZIF-72, theory as well as experiment, agree that the electronic permittivity obeys Clausius–Mossotti relation, while the phonon contribution disobeys the continuum mixing rules, even after considering rigorous analytical bounds. The physical origin is of a topological nature, but not chemical: framework connectivity partitions the cell into local regions and depolarize the large pore cavities, hence localizing the polar phonon eigenvectors and weakening the mode-effective Born charges. The macroscopic permittivity is therefore set by eigenvector coherence, not by dipoles.

Our findings could impact the emergent fields of MOF-based triboelectric nanogenerators and tunable dielectrics. For example, the ZIF-72 based triboelectrics substantially outperform the ZIF-71 counterpart due to the higher permittivity of the denser framework [14]. The underpinning mechanism can now be explained by our results associated with phonon localization, revealing new insights and offering powerful design guidelines. Likewise, as low-$\kappa$ MOF films are already under development for on-chip interconnects [44-46], an improved understanding on topology-driven phonon contribution could trigger exciting new developments. Finally, it is anticipated that the topology-controlled phonon dielectric response omnipresent in MOF materials should extend to other families of open-framework structures, encompassing covalent organic frameworks (COFs), hydrogen-bonded organic frameworks (HOFs), porous coordination polymers (PCPs), and other hybrid framework solids.

## ACKNOWLEDGMENTS

The authors acknowledge funding from the UKRI Engineering and Physical Sciences Research Council (EPSRC) award (TEGMOF EP/Z534146/1). Computational resources were provided by the University of Oxford Advanced Research Computing (ARC) facility; the ARCHER2 UK National Supercomputing Service, accessed through membership of the Materials Chemistry Consortium, which is funded by EPSRC grant EP/R029431/1; and the Science and Technology Facilities Council (STFC) Scientific Computing Department's SCARF high-performance computing cluster. The authors gratefully acknowledge Dr Svemir Rudić of UKRI STFC for facilitating access to the SCARF computing facility at Harwell. We acknowledge the Diamond Light Source for the award of beamtime SM43624, and for the technical support kindly offered by Dr. Hendrik Vondracek and Dr. Gianfelice Cinque on Beamline B22 MIRIAM.

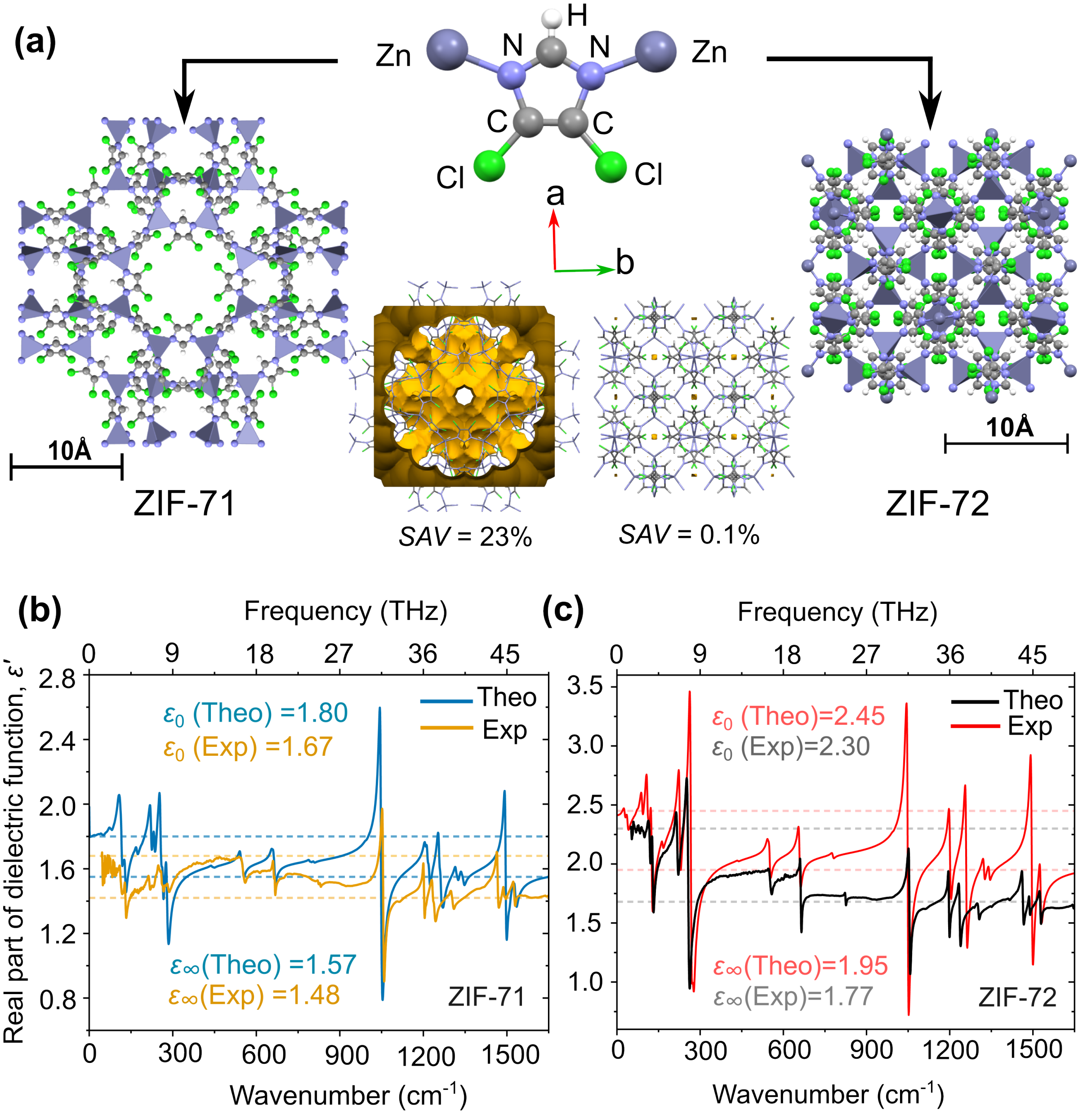

**FIG. 1.** (a) Crystal structures of ZIF-71 (RHO, porous) and ZIF-72 (LCS, dense), both built from the same Zn(II)–4,5-dichloroimidazolate. (b,c) Real part of the dielectric function $\mathrm{Re}[\varepsilon(\omega)] \equiv \varepsilon'$ for (b) ZIF-71 and (c) ZIF-72. Computed by DFT (Theo, black and blue) and measured by synchrotron infrared reflectance at the B22 MIRIAM beamline of the Diamond Light Source (Exp, red and orange). Horizontal lines mark $\varepsilon_0$ and $\varepsilon_\infty$.

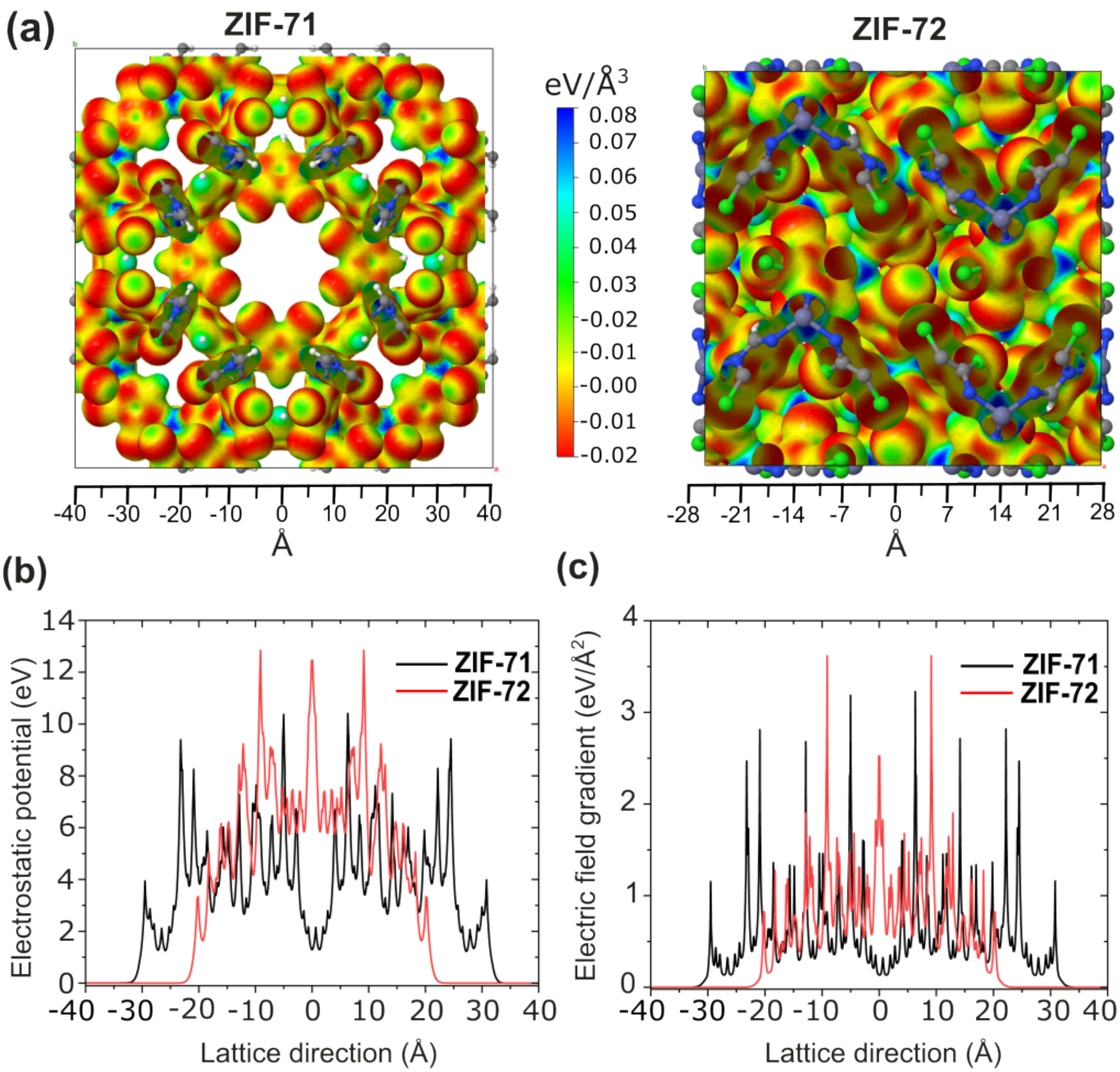

**FIG. 2.** 3D electrostatic potential ***V***(r) of (a) ZIF-71 (RHO) and ZIF-72 (LCS). (b) Planar-averaged potential *V(z)* and (c) the corresponding field gradient d*E*/d*z* along the cubic direction, showing the depolarization plateaus over the RHO cavities of ZIF-71 (at origin) and the continuous gradients across the dense unit cell of ZIF-72. Both frameworks are shown side by side in panel (a), note the different scales used for each unit cell.

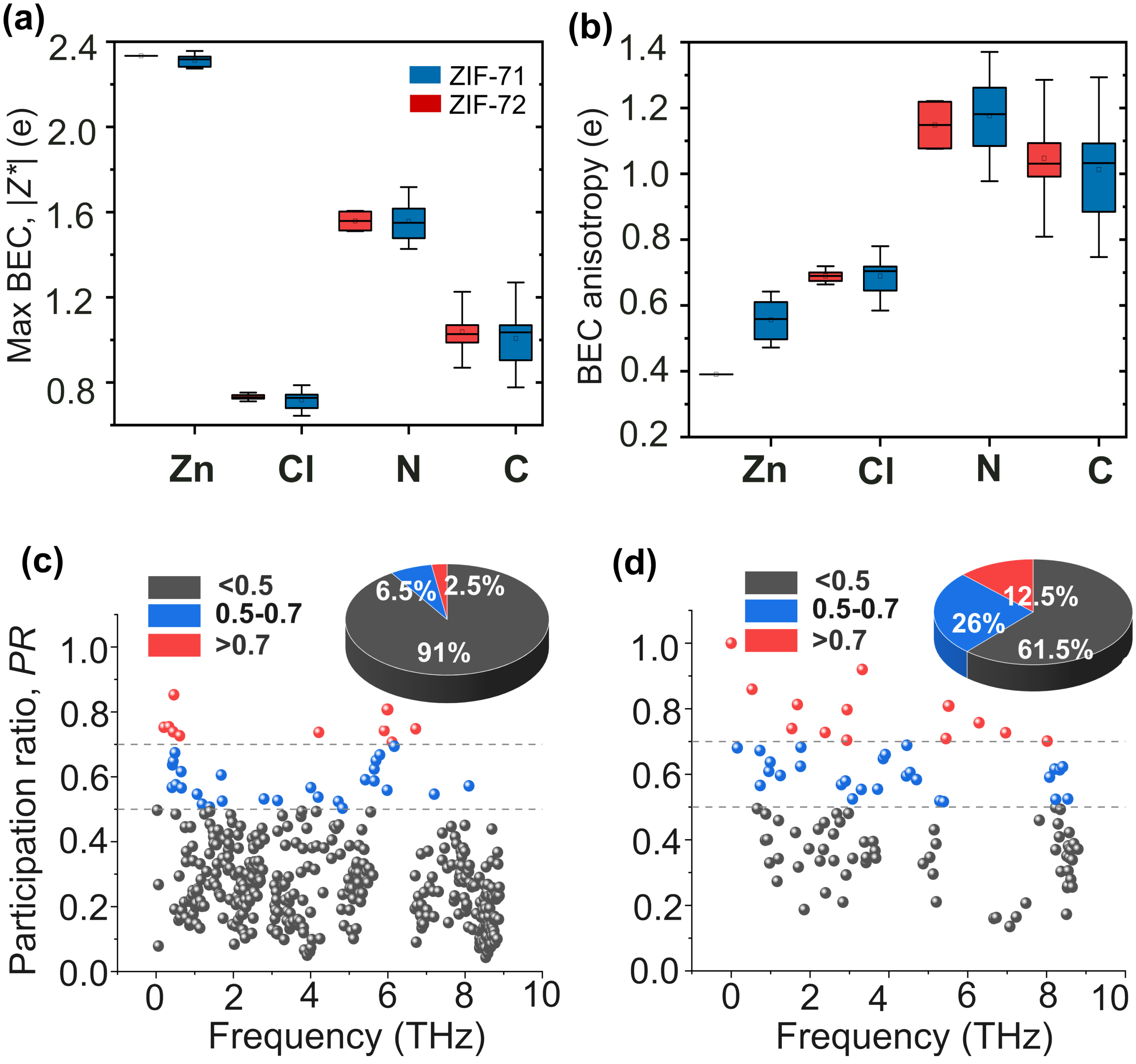

**FIG. 3.** (a) Box plot of the magnitudes of the principal Born-effective-charge tensor $|Z^*_{\max}|$ for each atomic species in ZIF-71 (blue) and ZIF-72 (red) and their medians equivalent. (b) Tensor anisotropy $|Z^*_{\max}| - |Z^*_{\min}|$ for each species. Participation ratio $PR$ vs phonon frequency for (c) ZIF-71 and (d) ZIF-72 in the phonon window 0–10 THz. Pie insets show the localized ($PR < 0.5$), intermediate ($0.5 \leq PR < 0.7$), and delocalized ($PR \geq 0.7$) fractions.

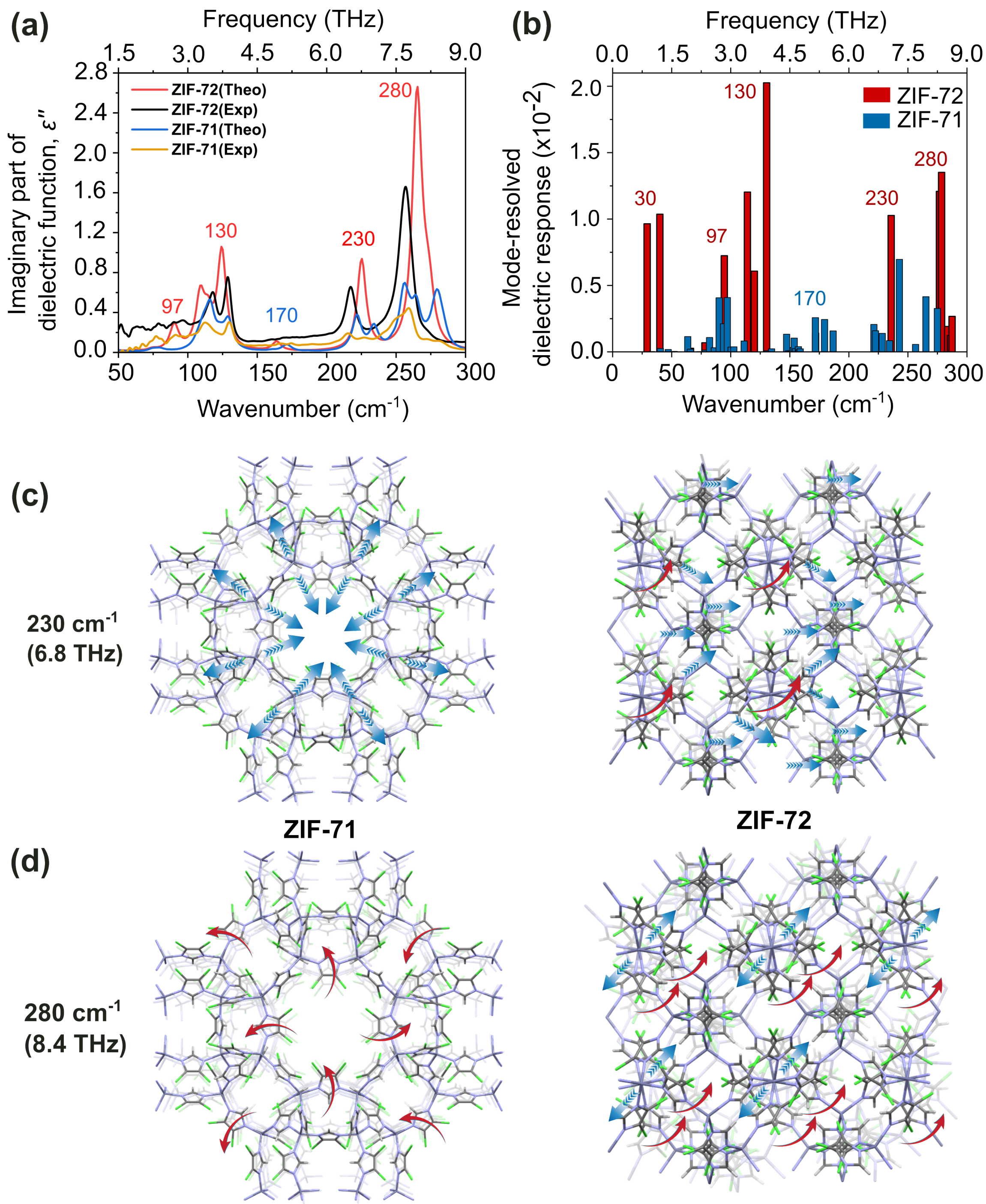

**FIG. 4.** (a) Imaginary part of the dielectric function, $\mathrm{Im}[\varepsilon(\omega)] \equiv \varepsilon''$ for ZIF-71 and ZIF-72, theoretical (DFT) and experimental. (b) Mode-resolved phonon contributions $\Delta\varepsilon_k$ for ZIF-71 (blue) and ZIF-72 (red); (c,d) Eigenvector displacements of the two highest-$\Delta\varepsilon_k$ modes of ZIF-72 at (c) 230 cm⁻¹ and (d) 280 cm⁻¹, with the corresponding modes of ZIF-71 at the same frequencies. Blue arrows indicate translational atomic displacements, and red arrows indicate rotation motions; video clips of the THz modes are provided as Movies S1–S4.

## REFERENCES

[1] J. K. Nayak, P. Roy Chaudhuri, S. Ratha, and M. R. Sahoo, J. Electromagn. Waves Appl. **37**, 282 (2023).
[2] D. A. G. Bruggeman, Ann. Phys. (Berlin) **416**, 636 (1935).
[3] G. J. Maxwell and B. Garnett, Philos. Trans. R. Soc. B. **203**, 385 (1904).
[4] A. H. Sihvola, *Electromagnetic mixing formulas and applications* (IET, 1999), 47.
[5] Z. Hashin and S. Shtrikman, J. Appl. Phys. **33**, 3125 (1962).
[6] K. Zagorodniy, G. Seifert, and H. Hermann, Appl. Phys. Lett. **97** (2010).
[7] A. M. Campos, J. Torres, and J. J. Giraldo, Surf. Rev. Lett. **09**, 1631 (2002).
[8] M. R. Ryder, L. Dona, J. G. Vitillo, and B. Civalleri, ChemPlusChem **83**, 308 (2018).
[9] M. R. Ryder *et al.*, Phys. Rev. Lett. **118**, 255502 (2017).
[10] A. I. Inamdar, S. Kamal, M. Usman, M.-H. Chiang, and K.-L. Lu, Coord. Chem. Rev. **502**, 215596 (2024).
[11] M. R. Ryder, B. Civalleri, T. D. Bennett, S. Henke, S. Rudić, G. Cinque, F. Fernandez-Alonso, and J.-C. Tan, Phys. Rev. Lett. **113**, 215502 (2014).
[12] O. M. Yaghi, M. O'Keeffe, N. W. Ockwig, H. K. Chae, M. Eddaoudi, and J. Kim, Nature **423**, 705 (2003).
[13] R. Banerjee, A. Phan, B. Wang, C. Knobler, H. Furukawa, M. O'Keeffe, and O. M. Yaghi, Science **319**, 939 (2008).
[14] J. Ye and J.-C. Tan, Nano Energy **114**, 108687 (2023).
[15] I. Sevostianov, S. G. Mogilevskaya, and V. I. Kushch, Int. J. Eng. Sci. **140**, 35 (2019).
[16] T. Wu, J. Zhang, C. Zhou, L. Wang, X. Bu, and P. Feng, J. Am. Chem. Soc. **131**, 6111 (2009).
[17] M. Ferrero, M. Rérat, R. Orlando, and R. Dovesi, J. Chem. Phys. **128** (2008).
[18] A. Erba *et al.*, J. Chem. Theory Comput. **19**, 6891 (2023).
[19] M. Ferrero, M. Rérat, R. Orlando, and R. Dovesi, J. Comput. Chem. **29**, 1450 (2008).
[20] F. Pascale, C. M. Zicovich-Wilson, F. López Gejo, B. Civalleri, R. Orlando, and R. Dovesi, J. Comput. Chem. **25**, 888 (2004).
[21] C. M. Zicovich-Wilson, F. Pascale, C. Roetti, V. R. Saunders, R. Orlando, and R. Dovesi, J. Comput. Chem. **25**, 1873 (2004).
[22] G. Rauhut and P. Pulay, J. Phys. Chem. **99**, 3093 (1995).
[23] A. S. Babal *et al.*, J. Phys. Chem. C **123**, 29427 (2019).
[24] K. Titov, Z. Zeng, M. R. Ryder, A. K. Chaudhari, B. Civalleri, C. S. Kelley, M. D. Frogley, G. Cinque, and J. C. Tan, J. Phys. Chem. Lett. **8**, 5035 (2017).
[25] A. F. Moslein and J. C. Tan, J. Phys. Chem. Lett. **13**, 2838 (2022).
[26] M. R. Ryder *et al.*, J. Phys. Chem. Lett. **9**, 2678 (2018).
[27] A. Phan, C. J. Doonan, F. J. Uribe-Romo, C. B. Knobler, M. O’Keeffe, and O. M. Yaghi, Acc. Chem. Res. **43**, 58 (2009).
[28] V. Lucarini, K.-E. Peiponen, J. J. Saarinen, and E. M. Vartiainen, *Kramers-Kronig relations in optical materials research* (Springer, 2005).
[29] X. Gonze and C. Lee, Phys. Rev. B **55**, 10355 (1997).
[30] R. Resta and D. Vanderbilt, in *Physics of ferroelectrics: a modern perspective* (Springer, 2007), pp. 31.
[31] S. Horike, D. Umeyama, and S. Kitagawa, Acc. Chem. Res. **46**, 2376 (2013).
[32] N. Bonanos, B. Steele, and E. Butler, Impedance spectroscopy, 205 (2005).
[33] J. D. Jackson, *Classical electrodynamics* (Wiley, New York, 1999), 3rd edn.
[34] R. G. Barrera and C. B. Duke, Phys. Rev. B **13**, 4477 (1976).

[35] C. Yuan, S. S. Sørensen, T. Du, Y. Song, and M. M. Smedskjaer, International Journal of Heat and Mass Transfer **233**, 126015 (2024).
[36] R. J. Bell and P. Dean, Discuss. Faraday Soc. **50**, 55 (1970).
[37] P. B. Allen and J. L. Feldman, Phys. Rev. B **48**, 12581 (1993).
[38] C.-Z. Wang, R. Yu, and H. Krakauer, Phys. Rev. B **54**, 11161 (1996).
[39] G. Marchese, F. Macheda, L. Binci, M. Calandra, P. Barone, and F. Mauri, Nat. Phys. **20**, 88 (2023).
[40] H. Schober and C. Oligschleger, Phys. Rev. B **53**, 11469 (1996).
[41] S. Burkov, B. Koltenbah, and L. Bruch, Phys. Rev. B **53**, 14179 (1996).
[42] L. Wang, A. Ninarello, P. Guan, L. Berthier, G. Szamel, and E. Flenner, Nat. Commun. **10**, 26 (2019).
[43] L. Giacomazzi, N. S. Shcheblanov, M. E. Povarnitsyn, Y. Li, A. Mavrič, B. Zupančič, J. Grdadolnik, and A. Pasquarello, Phys. Rev. Mater. **7** (2023).
[44] M. Krishtab, I. Stassen, T. Stassin, A. J. Cruz, O. O. Okudur, S. Armini, C. Wilson, S. De Gendt, and R. Ameloot, Nat. Commun. **10**, 3729 (2019).
[45] M. Usman and K.-L. Lu, NPG Asia Mater. **8**, e333 (2016).
[46] A. S. Babal, B. E. Souza, A. F. Möslein, M. Gutiérrez, M. D. Frogley, and J.-C. Tan, ACS Appl. Electron. Mater. **3**, 1191 (2021).

# *Supplementary Materials*

## Topology-Controlled Phonon Dielectric Response Beyond Density Scaling in Metal-Organic Frameworks

Debayan Mondal[1], Jiahao Ye[1], Lorenzo Donà[2], Jin-Chong Tan[1*]

[1]Multifunctional Materials & Composites (MMC) Laboratory, Department of Engineering Science, University of Oxford, United Kingdom.

[2]Department of Chemistry, NIS and INSTM Reference Centre, University of Turin, Italy.

[*]*Corresponding author*: jin-chong.tan@eng.ox.ac.uk

## Table of Contents

# S1. Computational Methodology

## S1.1 General Setup

All calculations were performed using the all-electron periodic density functional theory (DFT) code CRYSTAL23 [1], employing the atom-centred Gaussian-type orbital (GTO) basis sets within a linear combination of atomic orbitals (LCAO) framework. The hybrid PBEsol0-3c functional [2], incorporating 25% Hartree–Fock exchange with the PBEsol correlation, combined with the composite 3c correction for basis set incompleteness and dispersion was employed for all production calculations, and HSEsol-3c for cross-validation. Initial structural models for ZIF-71 (CCDC refcode: GITVIP01, space group $Pm\overline{3}m$, No. 221) and ZIF-72 (CCDC refcode: GIZJUV, space group $Ia\overline{3}d$, No. 230) were obtained from the Cambridge Structural Database (CSD) [3,4].

## S1.2 Basis Set

The sol-def2-mSVP basis set was used for all elements (Zn, C, N, Cl, H) [2]. This is a solid-state adapted contraction of the Karlsruhe def2-mSVP basis, specifically re-optimised for periodic calculations in CRYSTAL23 to reduce linear dependence issues that arise with diffuse exponents in crystalline environments. The basis set provides double-$\zeta$ quality with polarisation functions for all atoms and includes a Stuttgart-type effective core potential (ECP28) for Zn, treating the $3s^2 3p^6 3d^{10} 4s^2$ electrons explicitly while replacing the 28 innermost core electrons with a pseudopotential. Full contraction details for each element are available in the CRYSTAL23 online basis set library [1].

## S1.3 Geometry Optimisation

Full geometry optimisations (both lattice parameters and atomic positions) were performed using analytical gradients, with the experimental space group symmetry preserved throughout *via* the KEEPSYMM keyword (applied for ZIF-72). Convergence criteria were set as follows: maximum force on atoms $< 1 \times 10^{-5}$ a.u., maximum atomic displacement $< 1 \times 10^{-4}$ a.u., and energy change between optimisation steps $< 1 \times 10^{-7}$ Hartree. The DIIS (Direct Inversion in the Iterative Subspace) extrapolation technique was employed to accelerate self-consistent field (SCF) convergence. For ZIF-71, the BIPOSIZE keyword was set to 42,000,000 to accommodate the large memory requirements of the bipolar expansion in the exchange integrals for the 816-atom primitive cell of ZIF-71, see Table S1.

## S1.4 Summary of Computational Parameters

**Table S1.** Summary of computational parameters for ZIF-71 and ZIF-72.

| Parameter | ZIF-71 (RHO) | ZIF-72 (LCS) |
|---|---|---|
| CCDC refcode | GITVIP01 | GIZJUV |
| Space group (No.) | $Pm\bar{3}m$ (221) | $Ia\bar{3}d$ (230) |
| Atoms in primitive cell | 816 | 408 |
| Lattice parameter $a$ (Å) | 28.813 | 19.654 |
| Density (g cm$^{-3}$) | 1.12 | 1.77 |
| Functional | PBEsol0-3c | PBEsol0-3c |
| Basis set | sol-def2-mSVP | sol-def2-mSVP |
| Dispersion correction | D3(BJ) + ABC | D3(BJ) + ABC |
| Monkhorst–Pack $k$-mesh | 2 × 2 × 2 | 2 × 2 × 2 |
| TOLINTEG | 8 8 7 7 28 | 7 7 7 7 25 |

## S1.5 Dielectric and Vibrational Property Calculations

Static and frequency-dependent dielectric tensors were evaluated analytically within the Coupled-Perturbed Hartree–Fock/Kohn–Sham (CPHF/CPKS) framework [5,6]. The CPHF/CPKS approach computes the first-order response of the crystalline wavefunction to a homogeneous static electric field perturbation, yielding the electronic (high-frequency) dielectric tensor $\varepsilon_\infty$ and the Born effective charge tensors $\boldsymbol{Z}^*_\mu$, where $\mu$ is an atom index labelling the atoms in the unit cell. Harmonic vibrational frequencies and eigenvectors at the $\Gamma$-points were computed *via* numerical differentiation of analytical energy gradients (FREQCALC), with the INTENS keyword to activate infrared (IR) intensity calculations and the INTCPHF block to use the coupled-perturbed analytical approach for Born charges.

## S2. Experimental Methodology

### S2.1 Materials and Synthesis of ZIF-71 and ZIF-72 Nanoparticles

All reagents were used as received: zinc acetate dihydrate [$Zn(OAc)_2 \cdot 2H_2O$], zinc oxide (ZnO), and 4,5-dichloroimidazole (H-dcIm) were obtained from Sigma-Aldrich. The syntheses used methanol as a solvent.

The porous RHO phase (ZIF-71) was crystallised by ambient-temperature coordination-driven self-assembly in solution [7]. Zinc acetate dihydrate (2.4 mmol) and H-dcIm (9.6 mmol), a Zn:linker molar ratio of 1:4, were dissolved separately in 15 mL of methanol and homogenised for 1 h. The solutions were combined and stirred at room temperature for 24 h, during which deprotonation of the imidazole N–H and tetrahedral Zn–N bridging drove precipitation of a microcrystalline white solid. The product was recovered by centrifugation (8000 rpm, 10 min), washed three times with methanol to displace residual linker, and dried in air overnight. The dense LCS phase (ZIF-72) was obtained from the same building unit through a solvent-free thermal route that favours the close-packed polymorph [8]. Zinc oxide (2 mmol) and H-dcIm (6 mmol), a Zn:linker ratio of 1:3, were ground together and sealed in a 50 mL borosilicate vessel, then held at 150 °C for 24 h. Under these solvent-free conditions, condensation of the $ZnN_4$ tetrahedra proceeds with elimination of water to yield the dense, low-porosity network with an LCS topology. The pale-yellow powder was washed repeatedly with methanol to remove any unreacted linkers, collected by centrifugation (10,000 rpm, 10 min) over three cycles, and dried overnight.

Phase purity and topological assignment were verified by powder X-ray diffraction (XRD) (Rigaku MiniFlex, Cu K$\alpha$, $\lambda$ = 1.541 Å) against simulated patterns from the deposited crystal structures (CCDC GITVIP01 for ZIF-71, GIZJUV for ZIF-72). The diagnostic low-angle reflections of the RHO lattice ($2\theta \approx 4.4°$ and 7.6°) and the higher-angle reflections of the LCS lattice ($2\theta \approx 12.7°$ and 16.9°) were reproduced with no detectable secondary phases. The accessible-porosity contrast was confirmed by $N_2$ physisorption at 77 K, giving BET surface areas of 874 $m^2\ g^{-1}$ for ZIF-71 and 5.6 $m^2\ g^{-1}$ for ZIF-72, consistent with the open RHO cavity network and the non-porous LCS framework. These same activated batches were pelletised for the synchrotron specular-reflectance measurements (§S2.3).

### S2.2 Synchrotron IR Specular Reflectance Setup

Specular infrared (IR) reflectance experiments were conducted at Beamline B22 MIRIAM of the Diamond Light Source synchrotron (Harwell Campus, Oxfordshire, UK). Measurements were carried out using a Bruker Vertex 80V FTIR interferometer equipped with a Pike Technologies VeeMAX II variable-angle specular reflectance accessory. Reflectance spectra were collected on the pressed-powder pellets of ZIF-71 and ZIF-72 (diameter 13 mm, thickness ~1 mm) prepared under a similar condition for both frameworks. Spectra were measured at an incidence angle of 30° from the surface normal, with a spectral resolution of 2 $cm^{-1}$ and 512 scans averaged per measurement. The sample chamber was evacuated to better than $10^{-5}$ mbar and held at room temperature (~21 °C).

Far-infrared (FIR or THz) measurements covered the spectral range 30–1000 $cm^{-1}$, while the mid-infrared (MIR) measurements covered the range of 600–4000 $cm^{-1}$. Background spectra were collected by measuring the specular reflectance from a polished aluminium mirror immediately before each FIR and MIR run. Reliable signal was obtained down to 40 $cm^{-1}$; below this cutoff signal-to-noise degrades sharply, and we therefore restrict the experimental extraction of $\varepsilon(\omega)$ to the range of 30–4000 $cm^{-1}$ (~1.0–120 THz). Full instrumentation details and the schematic of the optical arrangement follow Titov *et al*. (2017) (Supporting Information of [9], Section 1). Related synchrotron far-IR and terahertz studies of ZIF frameworks have established this specular-reflectance approach for probing MOF-type materials [10,11].

### S2.3 Pellet Preparation and Sample Handling

ZIF-71 and ZIF-72 powders were synthesised in accordance with the published procedures [3,4]. Pellets were pressed using a 13-mm diameter stainless-steel die under uniaxial compression of about 1 ton (~76 MPa) for 5 minutes at room temperature. This load was selected on the basis of preliminary tests showing that pellets pressed under this condition exhibit reproducible specular reflectance and minimal surface roughness, while preserving the crystalline structure of both frameworks (verified by XRD before and after pelletisation). Both frameworks were activated by heating under dynamic vacuum at 120 °C for 12 h prior to pelletisation, then transferred to the FTIR sample chamber under inert atmosphere to avoid moisture uptake. Powder XRD patterns collected on the activated samples and on the as-pressed pellets confirmed that no phase changes or amorphisation occurred during pelletisation.

### S2.4 Kramers–Kronig Transformation (KKT) Procedure

The complex dielectric function $\varepsilon(\omega)$ was extracted from the measured reflectance $R(\omega)$ *via* the Kramers–Kronig transformation (KKT). For external reflection at near-normal incidence, the phase change $\theta(\omega)$ of the reflected field is related to the reflectance by

$$\theta_0 = -\frac{2\omega_0}{\pi} P \int_0^{\infty} \frac{\frac{1}{2}\ln\left[\frac{R(\omega)}{R_\infty}\right]}{\omega^2 - {\omega_0}^2} \, d\omega \qquad \text{(S1)}$$

where $P$ denotes the Cauchy principal value of the integral. The complex refractive index, $\tilde{n} = n + i\kappa$, follows from

$$n = \frac{1-R}{1+R-2\sqrt{R}\cos\theta}, \quad \kappa = \frac{-2\sqrt{R}\sin\theta}{1+R-2\sqrt{R}\cos\theta} \tag{S2}$$

and the complex dielectric function from $\varepsilon' = n^2 - \kappa^2, \varepsilon'' = 2n\kappa$. The KKT was implemented in MATLAB following the algorithm of Lucarini *et al.* [12] with extensions for finite-range far-IR (FIR) and mid-IR (MIR) data developed by Titov *et al.* [9]. The FIR and MIR reflectance spectra were joined at 600 $cm^{-1}$ using a piecewise cubic Hermite interpolating polynomial (PCHIP) and extrapolated to $\omega \to 0$ by holding the reflectance constant at its value at the lowest measured wavenumber. The full KKT routine, including MATLAB source code, is provided in Titov *et al.* (2017) (Supporting Information of [9], Section 10) and was used unmodified for the present work.

### S2.5 Drude–Lorentz Fitting and $\varepsilon_0$ Extraction

The static dielectric constant $\varepsilon_\infty^{\text{exp}}$ was extracted from the Kramers–Kronig-derived real part (Re) of the experimentally determined Re[$\varepsilon(\omega)$] by fitting a multi-oscillator Drude–Lorentz model [13,14] of the form

$$\varepsilon(\omega) = \varepsilon_\infty^{\text{exp}} + \sum_k \frac{f_k \omega_k{}^2}{\omega_k{}^2 - \omega^2 + i\omega\gamma_k} \tag{S3}$$

with $\omega_k$, $f_k$, and $\gamma_k$ the resonance frequency, oscillator strength, and damping constant of the $k^{\text{th}}$ IR-active mode. The static dielectric constant value, $\varepsilon_0^{\text{exp}}$, is recovered as the $\omega \to 0$ limit, $\varepsilon_0^{\text{exp}} = \varepsilon_\infty^{\text{exp}} + \sum_k f_k$. The determined experimental values are:

ZIF-71: $\varepsilon_0^{\text{exp}} = 1.67, \varepsilon_\infty^{\text{exp}} = 1.48$;

ZIF-72: $\varepsilon_0^{\text{exp}} = 2.30, \varepsilon_\infty^{\text{exp}} = 1.77$.

## S3. Theoretical Dielectric Function

### S3.1 Theoretical Framework

The static dielectric tensor of an insulating crystal is the sum of an electronic (clamped-ion) contribution and a phonon contribution:

$$\varepsilon_{0,\alpha\beta} = \varepsilon_{\infty,\alpha\beta} + \Delta\varepsilon_{\alpha\beta}^{\text{phon}} \tag{S4}$$

where $\varepsilon_\infty$ is the high-frequency clamped-ion dielectric tensor computed from the CPHF/CPKS response, and $\Delta\varepsilon_{\text{phon}}$ is the phonon contribution arising from the displacement of nuclei under the applied field [15]. The phonon contribution decomposes into individual contributions from each infrared-active phonon mode $k$:

$$\Delta\varepsilon_{\alpha\beta}^{\text{phon}} = \sum_k f_{k,\alpha\beta}, \quad f_{k,\alpha\beta} = \frac{1}{\varepsilon_0\, \Omega\, \omega_k{}^2} Z_{k,\alpha} Z_{k,\beta} \tag{S5}$$

where $f_{k,\alpha\beta}$ is the dimensionless oscillator strength of mode $k$, Ω is the unit cell volume, $\omega_k$ is the harmonic frequency of the mode, and $\tilde{Z}_{k,\alpha}$ is the mode-effective Born charge vector defined below.

### S3.2 Oscillator Strength Formula

As implemented in CRYSTAL23, the oscillator strength for mode $k$ is computed in atomic units as [16,17]

$$f_{k,\alpha\beta} = \frac{1}{\Omega\, \omega_k{}^2} Z_{k,\alpha} Z_{k,\beta} \tag{S6}$$

where the mode-effective Born charge vector $\tilde{Z}_{k,\alpha}$ is the projection of the atomic Born effective charge tensors $\boldsymbol{Z}^*_{\mu,\alpha\beta}$ onto the mass-weighted phonon eigenvectors $\mathbf{e}_{k,\mu,\beta}$:

$$Z_{k,\alpha} = \sum_{\mu,\beta} \frac{\boldsymbol{Z}^*_{\mu,\alpha\beta}\, \mathbf{e}_{k,\mu,\beta}}{\sqrt{M_\mu}} \tag{S7}$$

where $\mu$ is an atom index labelling the atoms in the unit cell; $M_\mu$ is the mass of atom $\mu$; and $\mathbf{e}_{k,\mu,\beta}$ are the Cartesian components (direction $\beta$) of the mass-weighted eigenvector of phonon mode $k$.

### S3.3 Mode-Effective Born Charges

Modes with large $|\tilde{Z}_k|$ (high mode-effective charge) and low frequency $\omega_k$ contribute disproportionately to the phonon-mediated dielectric constant, because $f_k$ scales as $|\tilde{Z}_k|^2 / \omega_k^2$. This is the formal basis for the localisation–dielectric link discussed in the main manuscript: a delocalised low-frequency phonon combined the coherently aligned atomic dipole contributions across many linkers, producing a large $|\tilde{Z}_k|$; a localised high-frequency phonon samples only one or two linkers, bounding $|\tilde{Z}_k|$ to that single dipole magnitude.

### S3.4 Drude–Lorentz Reconstruction of *ε*(*ω*)

The frequency-dependent dielectric function $\varepsilon(\omega)$ plotted in Fig. 1 of the main manuscript was reconstructed from the computed $\Gamma$-point phonon data using a classical damped harmonic oscillator (Drude–Lorentz) model with uniform Lorentzian broadening $\gamma_k$ = 4 cm$^{-1}$. In the static limit $\omega \rightarrow 0$, this reduces to $\varepsilon_0 = \varepsilon_\infty + \sum_k f_k$, recovering the full static dielectric constant used in the main manuscript.

# S4. Born Effective Charge Analysis

## S4.1 Definition and Physical Significance

The Born effective charge (BEC) tensor $\boldsymbol{Z}^*_{\mu,\alpha\beta}$ of atom $\mu$ is defined as the proportionality coefficient relating the change in macroscopic polarisation $P_\beta$ along direction $\beta$ to a collective sublattice displacement $v_{\mu,\alpha}$ of atoms $\mu$ along direction $\alpha$, under the condition of zero macroscopic electric field [18–21]. Its principal values $|Z^*_{max}|$ determine the magnitude of the dipole induced by atomic motion, while its anisotropy $Z^*_{max} - Z^*_{min}$ measures the directional character of the response.

## S4.2 Acoustic Sum Rule

Charge neutrality requires the sum of all atomic Born effective charges to vanish: $\sum_{\mu,\beta} \boldsymbol{Z}^*_{\mu,\beta} = 0$. This is the acoustic sum rule (ASR), guaranteeing that uniform translation of the entire crystal produces no macroscopic polarisation. Numerical violations of the ASR in CRYSTAL23 are typically below $10^{-3}$ *e* per atom, at which level we apply a uniform redistribution correction to enforce the rule exactly.

## S4.3 Computational Method

Born effective charge tensors were computed using the analytical Coupled-Perturbed Hartree–Fock/Kohn–Sham (CPHF/CPKS) approach in CRYSTAL23 [5,6] *via* the FREQCALC + INTENS + INTCPHF keyword combination. Principal values of $Z^*_{max}$, $Z^*_{med}$, $Z^*_{min}$ were obtained by diagonalisation of the symmetric part of the tensor $\boldsymbol{Z}^*_{\mu}$.

# S5. Phonon Participation Ratio Analysis

## S5.1 Definition and Computational Procedure

The participation ratio (*PR*) of a phonon mode *s* of frequency $\omega_s$ measures the fraction of atoms in the unit cell that participate meaningfully in that vibration: [22,23]

$$PR_k = \frac{\left(\sum_\mu A_{\mu,k}\right)^2}{\left[N \sum_\mu A_{\mu,k}{}^2\right]} \tag{S8}$$

where $A_{\mu,k} = \left|e_{\mu,k}\right|^2$ is the squared atomic displacement amplitude. *PR* ranges from $1/N$ (single-atom localised mode) to 1 (fully delocalised collective mode). Modes are classified as localised ($PR < 0.5$), intermediate ($0.5 \leq PR < 0.7$), or delocalised ($PR \geq 0.7$), following the standard convention [22,23].

## S5.2 PR Results for ZIF-71 and ZIF-72 (Table S2)

**Table S2.** Phonon participation ratio statistics for ZIF-71 and ZIF-72 within the lattice-polar window (0–10 THz). High-frequency intramolecular vibrations above 10 THz (> 333 $cm^{-1}$), which are intrinsically localised and contribute negligibly to $\Delta\varepsilon_{\text{phon}}$, are excluded.

| Property | ZIF-71 (RHO) | ZIF-72 (LCS) |
|---|---|---|
| Spectral window analysed | 0–10 THz (lattice-polar) | 0–10 THz (lattice-polar) |
| Localised modes (*PR* < 0.5) | 91.0% | 61.5% |
| Intermediate (0.5 ≤ *PR* < 0.7) | 6.5% | 26.0% |
| Delocalised (*PR* ≥ 0.7) | 2.5% | 12.5% |

### S5.3 Cutoff Robustness and Spectral Window

The *PR* statistics reported above are restricted to the lattice-polar window 0–10 THz, which contains the IR-active phonons that are responsible for the phonon-mediated dielectric weight (Fig. 3(c,d) of the main manuscript uses this same window on the THz axis). High-frequency intramolecular vibrations above 10 THz, including C–H stretches and ring deformations, are intrinsically localised regardless of framework topology and contribute negligibly to $\Delta\varepsilon_{\text{phon}}$ because of the $1/\omega^2$ weighting in the phonon resolved sum rule. Including or excluding these modes from the localised/delocalised classification does not change the qualitative contrast between the two frameworks reported in Table S2 but excluding them sharpens the *PR* contrast in the dielectrically relevant window. We further recomputed the fractions of localized (L), intermediate (I), and delocalized (D) states using two alternative participation-ratio (*PR*) thresholds: localized for *PR* < 0.5, intermediate for 0.5 ≤ *PR* ≤ 0.7, and delocalized for *PR* > 0.7. The qualitative result — ZIF-71 has substantially more localised modes than ZIF-72, and ZIF-72 retains a non-negligible delocalised fraction, holds across all three threshold pairs (not shown). The conclusions of the main manuscript are therefore robust to ±0.1 variation of the *PR* cutoffs and to the choice of frequency window within the lattice-polar (phonon) range.

## S6. Continuum Effective-Medium Benchmarks

### S6.1 Clausius–Mossotti Test (Table S3)

The Clausius–Mossotti relation links the static dielectric constant $\varepsilon$ of a dielectric to the molecular polarizability $\alpha$ and the number density $N$ of polarisable units through $\frac{\varepsilon-1}{\varepsilon+2}=\frac{N\alpha}{3\varepsilon_0}$. For two materials sharing the same molecular polarisability $\alpha$ per repeating unit — the case for an isochemical pair, such as ZIF-71 and ZIF-72 — this quantity is therefore expected to scale linearly with the mass density $\rho$, since $N \propto \rho$. The optical analogue, the Lorenz–Lorenz relation, has the same algebraic form but is written for the refractive index, $\frac{n^2-1}{n^2+2} = \frac{N\alpha}{3\varepsilon_0}$; using $\varepsilon_\infty = n^2$ this links the high-frequency dielectric channel directly to density. The dimensionless quantity that diagnoses simple density dilution of the dielectric is therefore,

$$\mathbb{C}(\varepsilon) = \frac{\varepsilon - 1}{\varepsilon + 2} \tag{S9}$$

which should scale linearly with mass density $\rho$ for an isochemical pair. For the (ZIF-71, ZIF-72) pair the density ratio was found to be $\rho_{\mathrm{ZIF-71}} / \rho_{\mathrm{ZIF-72}} = 1.12 / 1.77 = 0.633$. Table S3 reports the values for comparison.

**Table S3.** Clausius–Mossotti consistency check. The electronic-channel ratio $\mathbb{C}(\varepsilon_\infty)$ matches the density ratio 0.633 within 5–7%, confirming density dilution of the electronic dielectric channel. The ratio from phonon-contribution lies 27% (theory) and 43% (experiment) below the density ratio, identifying the phonon-mediated counterpart that no continuum effective-medium theory can reproduce.

| **Quantity** | **ZIF-71** | **ZIF-72** | **Ratio (ZIF-71/ZIF-72)** |
|---|---|---|---|
| Density $\rho$ (g cm$^{-3}$) | 1.12 | 1.77 | **0.633** |
| **Clausius–Mossotti quantity $\mathbb{C}(\varepsilon_\infty)$; electronic contributions** | | | |
| Theory (CPHF/KS) | 0.160 | 0.241 | 0.664 |
| Experiment (synchrotron IR + KKT) | 0.138 | 0.204 | 0.675 |
| **Phonon contribution ratio $\Delta\varepsilon_{\mathrm{phon}}$(ZIF-71) / $\Delta\varepsilon_{\mathrm{phon}}$ (ZIF-72)** | | | |
| Theory | 0.230 | 0.500 | **0.460** |
| Experiment | 0.190 | 0.530 | **0.358** |

## S6.2 Continuum Mixing Family: Maxwell–Garnett, Bruggeman, Hashin–Shtrikman

To establish that the residual from phonon-contribution lies outside the predictive reach of continuum effective-medium theory in general, and not merely outside Maxwell–Garnett, we evaluated $\varepsilon_0$ and $\Delta\varepsilon_{\text{phon}}$ of ZIF-71 from a representative family of two-phase isotropic continuum mixing rules [24], treating ZIF-72 as the dielectric host (porosity ≈ 0%) and ZIF-71 as the host plus vacuum inclusions at volume fraction $\varphi$ = 0.23 (the solvent-accessible volume of the RHO topology). The continuum mixing rules are:

**(i) Clausius–Mossotti dilution limit (spherical Maxwell–Garnett with the framework as inclusions in a vacuum matrix):**

$$\frac{\varepsilon_{\text{eff}} - 1}{\varepsilon_{\text{eff}} + 2} = (1 - \varphi) \cdot \frac{\varepsilon_{\text{host}} - 1}{\varepsilon_{\text{host}} + 2}$$

**(ii) Bruggeman symmetric effective-medium theory (EMT):**

$$(1 - \varphi) \cdot \frac{\varepsilon_{\text{host}} - \varepsilon_{\text{eff}}}{\varepsilon_{\text{host}} + 2\varepsilon_{\text{eff}}} + \varphi \cdot \frac{1 - \varepsilon_{\text{eff}}}{1 + 2\varepsilon_{\text{eff}}} = 0$$

which treats the two phases symmetrically and is appropriate for percolating microstructures [25]. Here, $\varphi$ multiplies the vacuum ($\varepsilon$ = 1) term and $(1-\varphi)$ multiplies the framework term. So $\varphi$ = void fraction and $(1-\varphi)$ = solid/framework fraction, used consistently, $\varphi \to 0$ gives $\varepsilon_{\text{eff}} \to \varepsilon_{\text{host}}$ (fully dense) and $\varphi \to 1$ gives $\varepsilon_{eff} \to 1$ (all vacuum).

**(iii) Hashin–Shtrikman bounds:** rigorous upper and lower bounds on the effective permittivity of any isotropic two-phase mixture at fixed volume fraction [26]. For the present case the two bounds coincide with the two spherical Maxwell–Garnett limits: the lower bound is the framework treated as high-permittivity inclusions dispersed in a vacuum matrix (the Clausius–Mossotti / Lorentz–Lorenz dilution), and the upper bound is vacuum inclusions dispersed in the high-permittivity framework host. This construction is rigorous for the present pair precisely because the isochemical constraint pins the dense-phase permittivity to ZIF-72; the HS bounds then apply to every isotropic mixture interpolating between ZIF-72 and vacuum, and a value below the lower bound is excluded for any such mixture. Both bounds are presented in Table S4.

All four estimates are reported in Table S4. Across the entire family, the electronic contribution $\varepsilon_\infty$ is reproduced within 6–7% of the actual ZIF-71 value, consistent with continuum theory; the phonon counterpart $\Delta\varepsilon_{\text{phon}}$ is overestimated by 29–36% (theory) and 46–50% (experiment), with no mixing rule able of reproducing the actual phonon-mediated permittivity of ZIF-71.

## S6.3 Depolarization Factor Sweep and Extended EMT Family (Table S4)

Table S4 summarises the predictions of the four representative continuum mixing rules listed in §S6.2, against the actual computed and measured ZIF-71 values. Moreover, we included the spherical-Maxwell–Garnett result swept across the physically reasonable range of $0.20 \leq L \leq 0.50$, which spans ellipsoidal inclusions from oblate to prolate. The results show that the failure of continuum mixing is a shared limitation, not an artefact of any single mixing rule or parameter choice. We note that a Maxwell–Garnett estimate at a single fixed depolarization factor $L$ corresponds to aligned (anisotropic) inclusions, so the swept values may lie slightly outside the isotropic Hashin–Shtrikman bounds; over the full sweep the shortfall of the predicted $\Delta\varepsilon_{\mathrm{phon}}$ relative to the actual value reaches up to 38% (theory) and 52% (experiment). In every case the measured and computed $\Delta\varepsilon_{\mathrm{phon}}$ lie below the entire mixing rule family.

**Table S4.** Predictions employing the two-phase isotropic continuum mixing rules for ZIF-71 (host = ZIF-72, vacuum inclusions at $\varphi = 0.23$), against the 'actual' computed (DFT) and experimentally (Exp) measured ZIF-71 values. †: Spherical-inclusion case, equivalent to the Hashin–Shtrikman lower bound for framework-in-vacuum geometry, across the effective-medium theory, namely Maxwell–Garnett, Bruggeman, both Hashin–Shtrikman bounds, and the depolarization-factor sweep.

| **Continuum estimate** | **$\varepsilon_0$ (Theo)** | **$\Delta\varepsilon_{phon}$(Theo)** | **Theoretical gap [Continuum − Actual (DFT)]/Actual (DFT) %** | **$\Delta\varepsilon_{phon}$(Exp)** | **Exp gap [Continuum − Actual (Exp)]/Actual (Exp) %** |
|---|---|---|---|---|---|
| **Continuum-Mixing at $\varphi$ = 0.23 ($\varepsilon_{host}$ ZIF-72 = 2.45, DFT)** | | | | | |
| Maxwell–Garnett ($L$ = 1/3, framework-in-vacuum) † | 2.005 | 0.323 | 40.4% | 0.350 | 84.2% |
| Bruggeman symmetric | 2.048 | 0.351 | 52.6% | 0.375 | 97.3% |
| Hashin–Shtrikman lower bound | 2.005 | 0.323 | 40.4% | 0.350 | 84.2% |
| Hashin–Shtrikman upper bound | 2.057 | 0.356 | 54.8% | 0.380 | 100% |
| **Maxwell–Garnett with $L$ sweep** | | | | | |
| $L$ = 0.20 (oblate) | 2.083 | 0.369 | 60.4% | 0.392 | 106.3% |
| $L \approx$ 0.33 (spherical) | 2.064 | 0.360 | 56.5% | 0.383 | 101.6% |
| $L$ = 0.50 (prolate) | 2.018 | 0.337 | 46.5% | 0.361 | 90% |
| **Actual ZIF-71 (this work)*** | **1.800 (DFT)** | **0.230 (DFT)** | — | **0.190 (Exp)** | — |

*From Eqn. S4, $\Delta\varepsilon_{phon} = \varepsilon_0 - \varepsilon_\infty$

**Conclusion of the continuum-mixing benchmark.** The shortfall of continuum mixing for the phonon-mediated dielectric constant is therefore not a numerical artefact of any particular formulation, but a structural consequence of the continuum-mixing assumption itself: averaging the inhomogeneous local field of the RHO cavities into a single homogenised mixing rule cannot reproduce the actual mode-by-mode phonon response. This is the quantitative evidence that supports the central claim of the main manuscript, that is the phonon-mediated dielectric residual lies beyond the family of continuum mixing rules, across their full physically reasonable parameter range.

## S7. Electrostatic Potential Mapping

### S7.1 Methodology

The ground-state electrostatic potential $V(\mathbf{r})$ was extracted from the converged CRYSTAL23 SCF charge density using the ECHG / POTM keywords. The full three-dimensional potential was sampled on a regular grid of 0.1 Å spacing within the crystallographic primitive cell, sufficient to resolve the framework–cavity boundary in both the ZIF-71 and ZIF-72 structures. Electron density isosurfaces were computed at $\rho = 0.005\ e$ Å$^{-3}$ and the potential mapped onto these surfaces (main manuscript, Fig. 2a). The planar-averaged potential $V(z)$, field $E(z)$, and field gradient d$E$/d$z$ were obtained directly from the CRYSTAL23 POTC module (option ICA = 2), which evaluates the exact potential and its first and second derivatives averaged over each ($xy$) plane orthogonal to the $z$-axis (Note: these directions correspond to the $a$-, $b$-, $c$- crystallographic axes). By Poisson's relation the field gradient is the induced charge density, $\rho(z) = -\frac{1}{4\pi}\frac{\mathrm{d}E}{\mathrm{d}z}$, so the near-zero d$E$/d$z$ plateau over the RHO cavity reflects a charge-free, field-excluded interior.

## S8. Generality Test: A Second Isochemical Pair of ZIF-8(SOD)/ZIF-8(dia)

To establish that topological control of the phonon-mediated dielectric channel is a general property of porous-versus-dense framework polymorphism, and not a feature peculiar to the ZIF-71/ZIF-72 (RHO/LCS) pair, we repeat the full analysis on another entirely independent isochemical pair: zeolitic imidazolate framework ZIF-8, with Zn(2-methylimidazolate)$_2$ building block in its porous sodalite (SOD) polymorph and its dense diamondoid (dia) polymorph. These two polymorphs also share an identical metal centre, ligand, and coordination tetrahedron ($ZnN_4$), and differ only in their long-range connectivity. The same electronic-versus-phonon dichotomy is reproduced, indicating that the residual coming from phonon contribution is a general rule for open-framework structures prevalent in MOFs.

### S8.1 Computational Protocol (Identical Level of Theory) (Table S5)

All ZIF-8(SOD)/ZIF-8(dia) quantities were computed in CRYSTAL23 at exactly the same level of theory and with the same protocol as the ZIF-71/ZIF-72 benchmark of §S1–§S6: hybrid PBEsol0-3c with the sol-def2-mSVP basis set, an ECP28 effective core potential on Zn and full optimisation of lattice and atomic coordinates under the experimental space-group symmetry. The clamped-ion (electronic) dielectric tensor $\varepsilon_\infty$ and the Born effective charges were obtained from the CPHF/KS response; the phonon contribution $\Delta\varepsilon_{\text{phon}}$ was obtained from the $\Gamma$-point IR oscillator strengths (FREQCALC + INTENS + INTCPHF). For the anisotropic dia polymorph ($P2_1/c$) the isotropic orientational average of the principal dielectric-tensor components is used throughout.

### S8.2 Summary of Computational Parameters

**Table S5.** Summary of computational parameters for ZIF-8 with sodalite (SOD) topology and ZIF-8 with diamond (dia) topology.

| Parameter | ZIF-8(SOD) | ZIF-8(dia) |
|---|---|---|
| CCDC refcode | FAWCEN01 | OFERUN01 |
| Space group (No.) | $I\bar{4}3m$ (217) | $P2_1/c$ (14) |
| Lattice parameter (Å) | $a$ = 16.9122<br>$b$ = 16.9122<br>$c$ = 16.9122 | $a$ = 17.5545<br>$b$ = 7.73070<br>$c$ = 14.8036 |
| Density (g cm$^{-3}$) | 0.93 | 1.56 |
| Functional | PBEsol0-3c | PBEsol0-3c |
| Basis set | sol-def2-mSVP | sol-def2-mSVP |
| Dispersion correction | D3(BJ) + ABC | D3(BJ) + ABC |
| Monkhorst–Pack $k$-mesh | 2 × 2 × 2 | 2 × 2 × 2 |
| TOLINTEG | 8 8 7 7 28 | 8 8 7 7 28 |

The primary computed inputs are as follows:

**ZIF-8(SOD)** (solvent-accessible volume fraction $\varphi$ = 0.24): $\varepsilon_\infty$ = 1.649, $\Delta\varepsilon_{\text{phon}}$ = 0.238, $\varepsilon_0$ = 1.887.

**ZIF-8(dia)** (fully dense, $\varphi$ = 0): $\varepsilon_\infty$ = 2.229, $\Delta\varepsilon_{\text{phon}}$ = 0.497, $\varepsilon_0$ = 2.726.

The static permittivity therefore rises by 44% from the porous to the dense polymorph for this isochemical pair of structures, for which the enhancement is attributed predominantly by the phonon contribution ($\Delta\varepsilon_{\mathrm{phon}}$ more than doubled, i.e. 0.238 → 0.497).

### S8.3 Clausius–Mossotti Test (Table S6)

**Table S6.** Clausius–Mossotti consistency check for the isochemical ZIF-8(SOD)/ZIF-8(dia) pair. The electronic-channel ratio $\mathbb{C}(\boldsymbol{\varepsilon}_\infty)$ lies 3% above the density ratio 0.594, confirming density dilution of the electronic dielectric channel at the same level of agreement as the ZIF-71/ZIF-72 pair. The phonon-contribution ratio lies 19% below the density ratio — the same direction of lattice phonon suppression as in ZIF-71/ZIF-72 (27%, theory) but reduced in magnitude, consistent with the smaller cavity of the SOD topology in ZIF-8 (relative to RHO in ZIF-71).

| Quantity | ZIF-8(SOD) | ZIF-8(dia) | Ratio (SOD/dia) |
|---|---|---|---|
| Density $\rho$ (g cm$^{-3}$) | 0.93 | 1.565 | **0.595** |
| **Clausius–Mossotti quantity $\mathbb{C}(\varepsilon_\infty)$** | | | |
| Theory (CPHF/KS) | 0.178 | 0.291 | 0.612 |
| **Phonon contribution ratio $\Delta\varepsilon_{\mathrm{phon}}$(SOD)/ $\Delta\varepsilon_{\mathrm{phon}}$ (dia)** | | | |
| Theory | 0.238 | 0.497 | **0.479** |

The electronic component of the ZIF-8 pair therefore obeys density dilution to within 3% ($L(\varepsilon_\infty)$ ratio 0.612 against density ratio 0.594), reproducing the Clausius–Mossotti behaviour of the ZIF-71/ZIF-72 electronic contribution. The phonon counterpart does not: the ratio of $\Delta\varepsilon_{\mathrm{phon}}$(SOD)/ $\Delta\varepsilon_{\mathrm{phon}}$ (dia) = 0.479 lies 19% below the density ratio. The architectural residual identified in the main manuscript is thus reproduced in a chemically distinct framework family, with the same sign and the same mechanism linked to THz modes in phonons.

## S9. Continuum-Mixing Theory for ZIF-8(SOD)/ZIF-8(dia) (Table S7)

Next, we tested the ZIF-8 phonon-contribution residual against the same representative family of two-phase isotropic continuum-mixing rules used for ZIF-71/ZIF-72 in §S6, treating the dense ZIF-8(dia) polymorph as the dielectric host ($\varphi = 0$) and the porous ZIF-8(SOD) polymorph as host-plus-vacuum-inclusions at $\varphi = 0.24$, the solvent-accessible volume fraction of the SOD topology. As before, the lower Hashin–Shtrikman bound coincides numerically with the spherical Maxwell–Garnett ($L$ = 1/3) result, and we additionally sweep the depolarization factor across $0.20 \leq L \leq 0.50$.

**Table S7.** Predictions of the representative family of two-phase isotropic continuum-mixing rules for ZIF-8(SOD) (host = ZIF-8(dia), vacuum inclusions at $\varphi$ = 0.24; the dia polymorph is fully dense, $\varphi$ = 0), against the actual computed ZIF-8(SOD) values. †: spherical-inclusion case, equivalent to the Hashin–Shtrikman lower bound for vacuum-in-host geometry. The gap from phonon-contribution therefore exceeds the electronic gap by a factor of roughly 2.5, compared with a factor of roughly 5 in ZIF-71/ZIF-72, indicating that the field-exclusion effect identified in the RHO framework is present but weaker in the smaller-cage SOD topology; This is consistent with the residual scaling with cavity dimension, although two polymorph pairs cannot by themselves establish a quantitative scaling law; additional pairs will be required to test it.

| **Continuum estimate** | **$\varepsilon_0$ (theo)** | **$\Delta\varepsilon_{phon}$ (theo)** | **Gap (continuum prediction − actual)/actual %** |
|---|---|---|---|
| **Continuum-Mixing at $\varphi$ = 0.24 ($\varepsilon_{host}$ [ZIF-8(dia)] = 2.72, DFT)** | | | |
| Maxwell–Garnett ($L$ = 1/3, framework-in-vacuum) † | 2.142 | 0.299 | 25.6% |
| Bruggeman symmetric | 2.209 | 0.336 | 41.1% |
| Hashin–Shtrikman lower bound | 2.142 | 0.299 | 25.6% |
| Hashin–Shtrikman upper bound | 2.223 | 0.344 | 44.5% |
| **Maxwell–Garnett with $L$ sweep** | | | |
| $L$ = 0.20 (oblate) | 2.258 | 0.358 | 50.4% |
| $L$ = 0.33 (spherical) | 2.232 | 0.347 | 45.8% |
| $L$ = 0.50 (prolate) | 2.170 | 0.322 | 35.3% |
| **Actual ZIF-8(SOD) (this work, DFT)** | **1.887** | **0.238** | — |

## S10. Phonon Participation-Ratio Analysis of ZIF-8(SOD)/ZIF-8(dia) (Table S8)

**Table S8.** Phonon participation-ratio (*PR*) analysis of the isochemical ZIF-8(SOD)/ZIF-8(dia) pair. In the lattice-polar THz window the dense dia polymorph carries the more delocalized (collective) phonon character: its most-delocalized mode is fully extended and lies at far lower frequency than that of the porous SOD polymorph, and dia uniquely supports a fully delocalized collective mode below 1 THz. The collective weight where the $1/\omega^2$ dielectric weighting is largest, consistent with its enhanced phonon-mediated permittivity (§S8–§S9).

| **Localization category** | **ZIF-8 (DIA)** | **ZIF (SOD)** |
|---|---|---|
| Localized ($PR < 0.5$) | 35 (76.1%) | 130 (91.5%) |
| Intermediate ($0.5 \le PR < 0.7$) | 5 (10.9%) | 8 (5.6%) |
| **Delocalized ($PR \ge 0.7$)** | **6 (13.0%)** | **4 (2.8%)** |
| Total modes (≤ 10 THz) | 46 | 142 |
| Mean *PR* | 0.42 | 0.32 |

## References

[1] A. Erba, J. K. Desmarais, S. Casassa, B. Civalleri, L. Donà, I. J. Bush, B. Searle, L. Maschio, L. Edith-Daga, A. Cossard, C. Ribaldone, E. Ascrizzi, N. L. Marana, J. P. Conesa, and R. Dovesi, J. Chem. Theory Comput. **19**, 6891 (2023).

[2] L. Donà, J. G. Brandenburg, and B. Civalleri, J. Chem. Phys. **151**, 121101 (2019).

[3] T. Wu, X. Bu, R. Liu, Z. Lin, J. Zhang, and P. Feng, J. Am. Chem. Soc. **131**, 6111 (2009).

[4] R. Banerjee, A. Phan, B. Wang, C. Knobler, H. Furukawa, M. O'Keeffe, and O. M. Yaghi, Science **319**, 939 (2008).

[5] M. Ferrero, M. Rérat, R. Orlando, and R. Dovesi, J. Chem. Phys. **128**, 014110 (2008).

[6] M. Ferrero, M. Rérat, R. Orlando, and R. Dovesi, J. Comput. Chem. **29**, 1450 (2008).

[7] Y. Zhang, M. Gutiérrez, A. K. Chaudhari, and J.-C. Tan, ACS Appl. Mater. Interfaces **12**, 37477 (2020).

[8] M. Tu et al., Angew. Chem. Int. Ed. **60**, 7553 (2021).

[9] K. Titov, Z. Zeng, M. R. Ryder, A. K. Chaudhari, B. Civalleri, C. S. Kelley, M. D. Frogley, G. Cinque, and J.-C. Tan, J. Phys. Chem. Lett. **8**, 5035 (2017).

[10] M. R. Ryder et al., J. Phys. Chem. Lett. **9**, 2678 (2018).

[11] A. R. Babal, A. F. Möslein, M. Nyman, M. R. Ryder, M. D. Frogley, and J.-C. Tan, J. Phys. Chem. C **125**, 14568 (2021).

[12] V. Lucarini, J. J. Saarinen, K.-E. Peiponen, and E. M. Vartiainen, Kramers–Kronig Relations in Optical Materials Research (Springer, 2005).

[13] R. Dovesi et al., J. Phys. Chem. C **123**, 8336 (2019).

[14] M. De La Pierre, C. Carteret, R. Orlando, and R. Dovesi, J. Comput. Chem. **34**, 1476 (2013).

[15] X. Gonze and C. Lee, Phys. Rev. B **55**, 10355 (1997).

[16] R. Dovesi et al., J. Chem. Phys. **152**, 204111 (2020).

[17] F. Pascale, C. M. Zicovich-Wilson, F. López, B. Civalleri, R. Orlando, and R. Dovesi, J. Comput. Chem. **25**, 888 (2004).

[18] Ph. Ghosez, J.-P. Michenaud, and X. Gonze, Phys. Rev. B **58**, 6224 (1998).

[19] R. Resta, Rev. Mod. Phys. **66**, 899 (1994).

[20] R. D. King-Smith and D. Vanderbilt, Phys. Rev. B **47**, 1651 (1993).

[21] R. Resta, Rev. Mod. Phys. **82**, 1959 (2010).

[22] R. J. Bell and P. Dean, Discuss. Faraday Soc. **50**, 55 (1970).

[23] P. B. Allen and J. L. Feldman, Phys. Rev. B **48**, 12581 (1993).

[24] A. H. Sihvola, Electromagnetic Mixing Formulas and Applications (IEE Publishing, London, 1999).

[25] D. A. G. Bruggeman, Ann. Phys. (Leipzig) **416**, 636 (1935).

[26] Z. Hashin and S. Shtrikman, J. Appl. Phys. **33**, 3125 (1962).